\newcommand{\LiCuO}{LiCu$_2$O$_2$}

\documentclass[aps,prl,twocolumn,superscriptaddress]{revtex4}
\usepackage{graphicx}
\usepackage{bm}

\begin{document}

{\bf Masuda et al. reply:} In our original work\cite{original} we
reported the observation of an incommensurate ordered state in the
frustrated quasi-one-dimensional antiferromagnet  \LiCuO. The
Comment by Drechsler {\it et al.} challenges our conclusions
regarding the hierarchy of relevant exchange interactions in the
system and the microscopic origin of frustration. In
Ref.~\cite{original} we postulated a simple model that seemed to
explain the available data with only two AF exchange constants
$J_1>J_2>0$ (see inset in Fig.~1). Drechsler {\it et al.} point
out that structural arguments and LDA calculations \cite{NMR}
favor a totally different picture \footnote{Drechsler {\it et al.}
use a different notation, related to ours through $J_1\rightarrow
J_\mathrm{DC}$, $J_2\rightarrow J_1$ and $J_4 \rightarrow J_2$.}:
$J_4
> -J_2 > 0$ and $J_1\sim 0$.

A determination of exchange parameters from bulk data is
notoriously ambiguous. To resolve the controversy we have instead
recently performed 3-axis inelastic neutron scattering experiments
that probe the coupling constants directly \cite{elsewhere}.
Fig.~1 (symbols) shows the spin wave dispersion measured along the
$(0.5,k,0)$ reciprocal space rod at $T=1.7$~K. Additional data
(not shown) were taken along $(h,0.827,0)$ and reveal a sinusoidal
dispersion with maxima at integer $h$ values and a  bandwidth of
7.5~meV. The measured dispersion curves can be analyzed in the
framework of linear spin wave theory (SWT) \cite{SWT}. It can be
shown that in the generalized $J_1$-$J_2$-$J_4$ model with
inter-chain coupling $J_\bot$ there are {\it exactly two} sets of
SWT coupling constants that fit the data: (i) $J_1=105$~meV,
$J_2=34$~meV, $J_4=-2$~meV and $J_\bot=0.2$~meV and (ii)
$J_1=6.4$~meV, $J_2=-11.9$~meV, $J_4=7.4$~meV and
$J_\bot=1.8$~meV. In the energy range shown in Fig.~1, the spectra
claculated from these two models (solid line) are
indistinguishable . Solution (i) almost exactly corresponds to our
original $J_1$-$J_2$ model. Note, however, that the fitted
effective $J$'s are unrealistically large. While this may merely
reflect severe quantum renormalization corrections, the
alternative model (ii) appears to be a more likely candidate for
\LiCuO. It incorporates a ferromagnetic $J_2$ bond, just like the
LDA-based model of \cite{NMR}. However, it involves only {\it
weak} frustration and requires a strong AF $J_1$ bond, as
originally proposed in our work. In addition, the estimated
inter-chain coupling constant is smaller than the LDA result by
half an order of magnitude. These two discrepancies will have
opposite effects on the Curie-Weiss temperature, which could in
turn explain why the LDA-based model still yields reasonable
estimates of this quantity.

Trying to reconcile the result by Drechsler {\it et al.} with the
measured dispersion of spin waves, we note that {\it just the data
taken along} $(0.5,k,0)$ can be also perfectly reproduced by
$J_1=0$, $J_2=-10$~meV, $J_4=7$~meV and $J_\bot=8$~meV. This set
of parameters is at least qualitatively consistent with their
model. However, with these numbers SWT gives an $a$-axis bandwidth
of 13~meV, almost twice as large as observed. One possibility is
that Drechsler's model is actually correct, but SWT breaks down
{\it qualitatively}, and can not give correct excitation energies
in the entire Brillouin zone {\it even using some effective set of
renormalized coupling constants}. This intriguing possibility
deserves a closer theoretical investigation, but seems unlikely.
Indeed, in \LiCuO\ the suppression of $T_c$ is not too pronounced,
and a renormalized quasiclassical picture should work rather well.

\begin{figure}
\includegraphics[width=8cm]{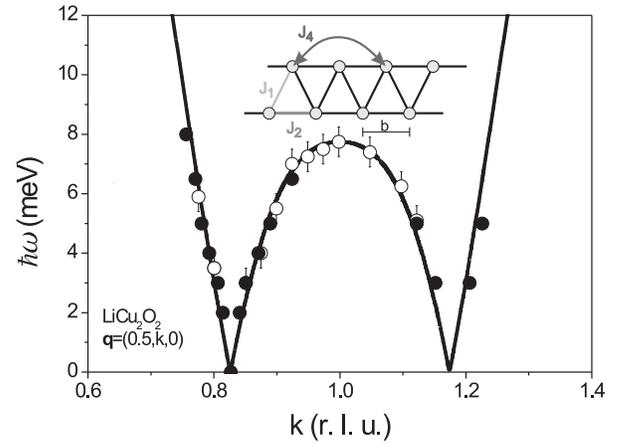}
\caption{Spin wave dispersion in \LiCuO\ measured using
constant-$E$  (solid symbols) or constant-$Q$ scans (open
symbols). Lines are as described in the text.}
\end{figure}

In summary, the frustration mechanism in \LiCuO\ is more complex
than we originally thought, and involves a {\it ferromagnetic}
$J_2$ bond. However, our present understanding of the inelastic
neutron scattering results suggests a strong ``rung'' interaction
$J_1$ and weak inter-chain coupling, in contradiction with the
model of Drechsler {\it et al.}  This work was partially supported
by the Civilian Research and Development Foundataion project
RU-P1-2599-04. Work at ORNL was supported by the U. S. Department
of Energy under Contract No. DE-AC05-00OR22725.

\vfill

T. Masuda$^{\dag}$, A. Zheludev$^{\dag}$,A. Bush$^{\sharp}$, M.
Markina$^{\flat}$, and A. Vasiliev$^{\flat}$\\
 $^{\dag}$~Condensed Matter Sciences Division, Oak Ridge National Laboratory,
Oak Ridge, TN 37831-6393, USA.\\
 $^{\sharp}$~Moscow Institute of
Radiotechnics, Electronics and Automation, Moscow 117464,
Russia.\\
 $^{\flat}$~Low Temperature Physics Department, Moscow State University,
Moscow 119992, Russia.\\

\end{document}